\documentclass[a4paper]{jpconf}
\usepackage{graphicx}
\begin{document}

\title{Cosmological constraints from supernova data set with corrected redshift}

\author{A. Feoli$^{1}$, L. Mancini$^{2,3,4}$, V. Rillo$^{1}$, M. Grasso$^{1}$}

\address{$^{1}$ Department of Engineering, University of Sannio, Piazza Roma 21, 82100 -- Benevento, Italy}
\address{$^{2}$ Max Planck Institute for Astronomy, K\"{o}nigstuhl 17, 69117 -- Heidelberg, Germany}
\address{$^{3}$ Department of Physics, University of Salerno, Italy}
\address{$^{4}$ International Institute for Advanced Scientific Studies, Vietri sul Mare (SA), Italy}

\ead{feoli@unisannio.it}

\begin{abstract}
Observations of distant type Ia supernovae (SNe Ia), used as
standard candles, support the notion that the Cosmos is filled
with a mysterious form of energy, the \emph{dark energy}. The
constraints on cosmological parameters derived from data of SNe Ia
and the measurements of the cosmic microwave background
anisotropies indicate that the dark energy amounts to
$\approx70$\% of all the energy contained in the Universe. In the
hypothesis of a flat Universe ($\Omega_{\mathrm{m}} +
\Omega_{\Lambda} =1$), we investigate if the dark energy is really
required in order to explain the SNe Ia experimental data, and, in
this case, how much of such unknown energy is actually deduced
from the analysis of these data and must be introduced in the
$\Lambda$CDM model of cosmology. In particular we are interested
in verifying if the Einstein--de Sitter model of the expanding
Universe is really to be ruled out. By using a fitting procedure
based on the Newton method search for a minimum, we reanalyzed the
``Union compilation'' reported by Kowalski et al. (2008) formed by
307 SNe, obtaining a very different estimate of the dark energy,
that is $\approx 60$\%. Furthermore, in order to balance the
correction of the apparent magnitude of SNe Ia, due to the
dilation or stretching of the corresponding light curve width, we
introduce a suitable modified redsfhit. Taking into account this
correction, we refitted the Union compilation dataset after a
selection cut. The main result that emerges from our analysis is
that the values of $\Omega_{\mathrm{m}}$ and $\Omega_{\Lambda}$
strongly depend on the fitting procedure and the selected sample.
In particular, the constraint we obtain on the mass density,
normalized by the critical mass density,  is $\Omega_{\mathrm{m}}
= 0.7$ for a sample of 252, and $\Omega_{\mathrm{m}}=1$ for a
sample of 242 SNe Ia respectively. The latter case does not imply
the existence of any additional form of dark energy.
\end{abstract}

\section{Introduction}
\label{Sec01}
It is well known that the  supernovae (SNe) Ia have a significant
intrinsic brightness due to their very homogenous initial
conditions, i.e. thermonuclear explosions of white dwarfs. For
this reason they are among the best standard candles providing
cosmological distance determination, with a precision of
approximately 8\%, and thus determination of cosmological
parameters \cite{perlmutterSchmidt2003}. The detection of SNe Ia
thus turns out to be essential in order to probe the cosmological
expansion (for a review, see ref. \cite{perlmutter2003}).

After the first systematic efforts to detect low--redshift
($z<0.1$) SNe Ia, carried out by the Cal\`{a}n/Tololo Supernova
Survey \cite{hamuy1993,hamuy1995,hamuy1996}, several international
teams, the Supernova Cosmological Project (SCP) and the High-Z SN
Search (HZSNS) performed extensive searches for SNe Ia at high
redshift (out to $z \leq 0.8$)
\cite{perlmutter1997,perlmutter1998,garnavich1998,schmidt1998,riess1998}.
In the wake of these successful experiments, a fit of the
magnitude-redshift data of 42 SNe Ia (at redshifts between 0.18
and 0.83), discovered by the Supernova Cosmology Project (SCP),
jointly with a set of 18 supernovae from the Cal\`{a}n/Tololo
Supernova Survey, was performed at redshifts below 0.1
\cite{perlmutter1999}. Practically, this study compared the
distance modulus of the measured SNe Ia with that which would be
expected for an Einstein--de Sitter flat Universe, leading to the
truly surprising discovery of a positive vacuum energy density
($\Omega_{\Lambda}>0$), i.e. the \emph{dark energy}, which opposes
the self--attraction of matter and causes the expansion of the
Universe to accelerate (for a review, see refs.
\cite{frieman2008,barris2010}). This result is consistent with the
full sky temperature maps based on seven years of data from WMAP,
which are well fit by a minimal six-parameter flat $\Lambda$CDM
model. In fact, by using the WMAP data in conjunction with baryon
acoustic oscillation data from the Sloan Digital Sky Survey and
priors on the Hubble constant $H_{0}$ from Hubble Space Telescope
observations, it was possible to estimate that
$\Omega_{\Lambda}=0.728_{-0.016}^{+0.015}$ \cite{jarosik2011}.

In more recent years, several large-scale ground-based projects
have been started
\cite{filippenko2001,nobili2005,copin2006,hamuy2006,law2009}, the
Hubble Space Telescope playing a key role in high-precision
optical and infrared follow-up of SNe discovered from the ground.
It was therefore possible to draw up new SN Ia compilations,
combining high-redshift ($z\sim0.5$) with low-redshift
($z\sim0.05$) samples, collected in different projects
\cite{tonry2003,knop2003,barris2004,riess2004,astier2006,wood-vasey2007,riess2007}.
However, the corresponding cosmological constraints were
consistent with a very wide range of dark energy models. This fact
is basically caused by statistical errors but, most importantly,
by systematic uncertainties as well. As a matter of fact, the
statistical uncertainties in measurements of dark energy
parameters are on the same order as systematic uncertainties.
Moreover, the different results significantly depend on which
light curve fitter has been used. This fact clearly emerges from
subsequent works
\cite{kowalski2008,holtzman2008,hicken2009,kessler2009}. It is
also remarkable to note that the constraints on the value of
$\Omega_{\mathrm{\Lambda}}$, which is the main goal of this
research, depend strongly on redshift of the SNe Ia. In fact, the
data at $z>1$ are not able to provide robust constraints on the
existence of the dark energy. Lastly, the physical nature of the
dark energy is at the moment lost in the meanders of theoretical
speculations \cite{copeland2006,ellis2008,buchert2008,durrer2008,linder2008,linder2009,%
mattsson2010,kamionkowski2011}.

In the present paper, we challenge the current standard model of
cosmology, the $\mathrm{\Lambda}$CDM model, which includes the
cosmological constant as an essential feature. In particular we
debate on the use of the stretch factor that intervenes in the SN
Ia light curves to modify the time axis around the peak of the
luminosity. Actually, without this empirical transformation, the
light curves of different SNe Ia are not congruent with each
other, but have a shape and a maximum that differ substantially
among the samples. However, we believe that any correction applied
on the apparent magnitude of each SNe Ia implies a correction on
its corresponding redshift, as we discuss in section 2. From this
perspective we have reanalysed the so--called ``Union
compilation'', which is a compilation of 414 SNe Ia (which reduces
to 307 after selection cuts) provided by ref. \cite{kowalski2008}
together with a framework to analyze datasets in a homogeneous
manner\footnote{Recently this analysis chain has been further
refined in ref. \cite{amanullah2010}, the ``Union2 sample''.}.

The paper is structured as follows. In \S \ref{Sec02} we introduce
the correction that we carried on the redshift of SNe. In \S
\ref{Sec03} we explain the fitting procedure and the selection
cuts on the original sample that have led to the compilation of
two subsets. The resulting parameters of the fits are reported
together with the Hubble diagrams. Finally, in \S \ref{Sec04} we
summarize and discuss the results.

\section{Apparent magnitudes and Redshift corrections}
\label{Sec02}
The apparent magnitudes of SNe measured by different groups at
different redshifts must be corrected, before putting them
together in the Hubble diagram, in such way that the absolute
magnitude can be considered the same constant for all the SNe
forming the sample. The photometric images are corrected for
galactic extinction, $K$--correction, etc., and finally undergo a
light-curve width-luminosity correction \cite{perlmutter1997}. The
light curve of each SN is compared with a template and the peak
apparent magnitude is corrected assuming a functional form
\cite{perlmutter1999}
%
\begin{equation}
m^{\mathrm{eff}} = m^{\mathrm{peak}} + \alpha(s - 1), %
\label{Eq_01}
\end{equation}
where $\alpha$ is a coefficient that will be determined by a
fitting algorithm, simultaneously with the other cosmological
parameters; $s$ is the so called ``stretch factor'' that suitably
stretches or contracts the time axis of each light curve around
the date of maximum light, affecting both the rising and declining
part of the light curve of each SN.

The stretch factor was introduced to reduce the intrinsic
dispersion in the absolute brightness of supernovae, but it really
acts as a physical mechanism and it was first considered due to a
local astrophysical effect (for example the temperature -
dependent variation in the opacity) by the authors in ref.
\cite{perlmutter1997}. Then, four years later, they same authors
admitted that: ``\textit{a single stretch factor varying the
timescale of the SNe Ia accounts very well for the restframe $B$
band light curve both before and after maximum light}'' but
``\textit{It is not understood from the current status of the
theory of SNe Ia whether this is a fortuitous coincidence or a
reflection of some physics timescale}'' \cite{goldhaber2001}.

The behavior of the SNe Ia light curves is usually interpreted as
a strong support for time dilation due to an expanding Universe.
However, non-standard cosmological models have been suggested as
alternative explanations for acceleration without dark energy. As
a matter of fact, if we use a model-independent method to compare
quantitatively the ability of some dark energy models and
non-standard types of cosmology (braneworld, $f (R)$, kinematic
models, etc.) to represent the Union2 sample data
\cite{amanullah2010}, none of the considered cosmological models
can be rigorously rejected on the basis of the current SN Ia data
\cite{benitez2012}.

Here we propose an alternative view of the stretch factor that we
interpret as a virtual displacement of the supernova from the
measured redshift to a new effective redshift. Actually the width
of the light curves is larger for SNe Ia at higher redshift
($z>0.1$) than it is for local ones. In fact, due to redshift, the
observed width evolves by a \emph{time-dilatator} factor $(1+z)$.
Still the same authors \cite{goldhaber2001} adjusted the SCP and
Cal\`{a}n/Tololo photometric data to the maximum intensity
$I_{\mathrm{max}}$ and scaled the time axis by a width factor $w =
s(1+z)$. Remarkably, with the application of this single stretch
timescale parameter, the SNe Ia data fell all on a common curve at
the level of the measurement uncertainty.

However, in our opinion, the correction of the apparent magnitude
should be balanced by a correspondent correction of the redshift.
In fact, since the transformation from the observer frame to the
SN rest frame is obtained by dividing the time interval by the
factor $w=s(1+z)$, this is equivalent to consider a time-dilatator
factor $(1 + z_{\mathrm{corr}})=s(1+z)$, and SNe shall not be
considered at redshift $z$ anymore, but at somewhat corrected
redshift $z_{\mathrm{corr}}$, such that
%
\begin{equation}
z_{\mathrm{corr}} = s + sz - 1. %
\label{Eq_02}
\end{equation}
In the light of our interpretation of the stretch factor, we think
that the SN Ia data have to be refitted by using the above
corrected redshift, which takes into account nothing else than the
correction of the apparent magnitude of SNe adducted by the
scientists of the SCP on the photometric data.

\section{Hubble diagram construction and cosmological parameter fitting}
\label{Sec03}
We considered the data reported in ref. \cite{kowalski2008}, even
if these authors did not strictly preserved the original meaning
of $s$. Actually, in the recent update of the SNe sample
\cite{amanullah2010} the meaning of the stretch factor was
completely lost because the new fitting algorithm SALT2 returns a
parameter $x_1$ that must be converted to the old $s$ using a
correlation formula that would become a source of other
statistical errors.

The effective magnitude of the $i$--th of the $N$ SNe forming our
modified sample is obtained from the formula:
%
\begin{equation}
m^{\mathrm{eff}}_i = m^{\mathrm{peak}}_i + \alpha(s_i - 1)- \beta c_i %
\label{Eq_03}
\end{equation}
where $c$ is the rest frame color at maximum as defined in ref.
\cite{kowalski2008} and $\beta$ is an additional parameter, which
has been determined by the fitting procedure too.

The $\chi^2$ corresponding to that of eq.(\ref{Eq_03}) is given as
%
\begin{equation}
\chi^2= \sum_{i=1}^N \frac{(m^{\mathrm{eff}}_i - 5 \log_{10} D_i -
\widetilde{M})^2}{(\sigma_{m \, i} + \alpha \, \sigma_{s \, i} -
\beta
\, \sigma_{c \, i})^2 + \sigma_{\mathrm{int}}^2 + \sigma_{v \, i}^2}, %
\label{Eq_04}
\end{equation}
where $D$ is the Hubble free luminosity distance
%
\begin{equation}
D =(1+z) \int_0^z \frac{dz^\prime}{\sqrt{\Omega_m (1+z^\prime)^3 + \Omega_\Lambda}},%
\label{Eq_05}
\end{equation}
$\widetilde{M}$ is the magnitude zero point offset
\begin{equation}
\widetilde{M} = M + 5 \log_{10} \frac{C}{H_0} + 25,%
\label{Eq_06}
\end{equation}
$C$ is the speed of light, and $M$ is the absolute magnitude which
is assumed to be constant after that the above mentioned
corrections of apparent magnitudes have been implemented.
$\sigma_{m}$, $\sigma_{s}$, $\sigma_{c}$ are the errors related to
the magnitude, stretch factor, and color factor, respectively. We
neglected the errors on $z$ and $z_{\mathrm{corr}}$. As in ref.
\cite{nesseris2005}, we considered an error due to a peculiar
velocity dispersion of $300$ km sec$^{-1}$, which can be written:
%
\begin{equation}
\sigma_{v \,i} = \frac{0.005}{\ln 10} \left(\frac{1}{1+z_i} +
\frac{1}{\sqrt{\Omega_{\mathrm{m}} (1+z_i)^3 +
\Omega_\Lambda}\int_0^{z_i}\frac{dz}{\sqrt{\Omega_{\mathrm{m}}
(1+z)^3 + \Omega_\Lambda}}} \right). %
\label{Eq_07}
\end{equation}
Furthermore, we supposed that our Universe is flat, so that
$\Omega_{\mathrm{m}} + \Omega_\Lambda =1$.

We found a minimum of our $\chi^2$ in eq.(\ref{Eq_04}) starting
from an intrinsic dispersion of the relation
$\sigma_{\mathrm{int}}=0.15$, and we updated this value refitting
until we have obtained a reduced $\chi^2_{\mathrm{red}} =
\chi^2/\mathrm{DOF} \simeq 1$. To this aim we used the ``Newton
method search for a minimum'' developed in Mathematica in a way
very similar to the algorithm adopted in ref. \cite{nesseris2005}.
The same results  can be equivalently obtained using iteratively
the ``NonlinearRegress'' functionality of Mathematica.

We applied this procedure to the standard sample of 307 SNe Ia
reported in ref. \cite{kowalski2008}. The resulting four
parameters ($\Omega_{\mathrm{m}}$, $\widetilde{M}$, $\alpha$,
$\beta$) of the fitting procedure are listed in Table
\ref{Table_1}, which clearly shows, in the first row, a difference
of more than $10\%$ in the estimate of $\Omega_{\mathrm{m}}$
between our procedure and the well known standard result.

Next, we corrected the measured redshift $z$ by eq.(\ref{Eq_02}),
and after this correction, we neglected the 35 following SNe
because their $z_{\mathrm{corr}}$ are now negative: 1993ag, 1993o,
1993h, 1992br, 1992bp, 1992bo, 1992bl, 1992aq, 1990af, 2001cn,
2000bh, 1993ac, 1994m, 1994t, 1995ak, 1996bo, 2000fa, 2000dk,
2000cn, 2000cf, 1999gd, 1999ek, 1999cc, 1998eg, 1998ef, 1998dx,
1998co, 1998ab, 1998v, 1997dg, 1997y, 1999bm, 1995ap, h283, n326.
\begin{table}
\caption{Fitting procedure results.}%
\begin{center}
\begin{tabular}{c c c c c c c c } 
\br
$z$ & $N$ & $\Omega_{\mathrm{m}}$ & $\widetilde{M}$ & $\alpha$ & $\beta$ & $\sigma_{\mathrm{int}}$ & $\chi_{\mathrm{red}}^{2}$\\
\mr
normal    & 307 & $0.393 \pm 0.046$ & $24.015 \pm 0.035$ & $0.81  \pm 0.12$ & $1.15 \pm 0.10$ & 0.25 & 1.056 \\
corrected & 252 & $0.70 \pm 0.22$   & $24.24  \pm 0.11$  & $11.01 \pm 0.41$ & $1.25 \pm 0.28$ & 0.32 & 1.005 \\
corrected & 242 & $1$               & $24.400 \pm 0.034$ & $10.57 \pm 0.33$ & $1.23 \pm 0.24$ & 0.20 & 1.075 \\
\br
\end{tabular}
\label{Table_1} %
\end{center}
\end{table}

Then, we considered the remaining  272 SNe, and we made a fitting
test without taking into account the measurement errors, and
imposing $\Omega_{\mathrm{m}}=1$. Actually, we did not want a
really blind analysis of the data, but we wanted to test if the
Einstein-de Sitter model, without any cosmological constant,
should really be excluded from the cosmological theory.
Considering the parameters resulting from the test fit, 20  SNe
were removed from the 272 SN sample having the highest residuals
with respect to the Einstein-de Sitter model; they are: 1992ag,
1992ae, 1992p, 1990o, 1999gp, 1994s, 1999aa, 1999ao, 1999aw,
1996j, 1996cl, 2001jn, 2001jf, 040mb, 04Sas, 2003dy, e138, g142,
k396, p534. After this selection cut, we applied our fitting
procedure considering also the measurement errors and the
corresponding results are listed in the second line of Table 1. With respect to the
307 SN data set, we obtained a completely different estimation of
the mass density, i.e. $\Omega_{\mathrm{m}} = 0.7$.

\begin{figure}%
\begin{center}
\includegraphics{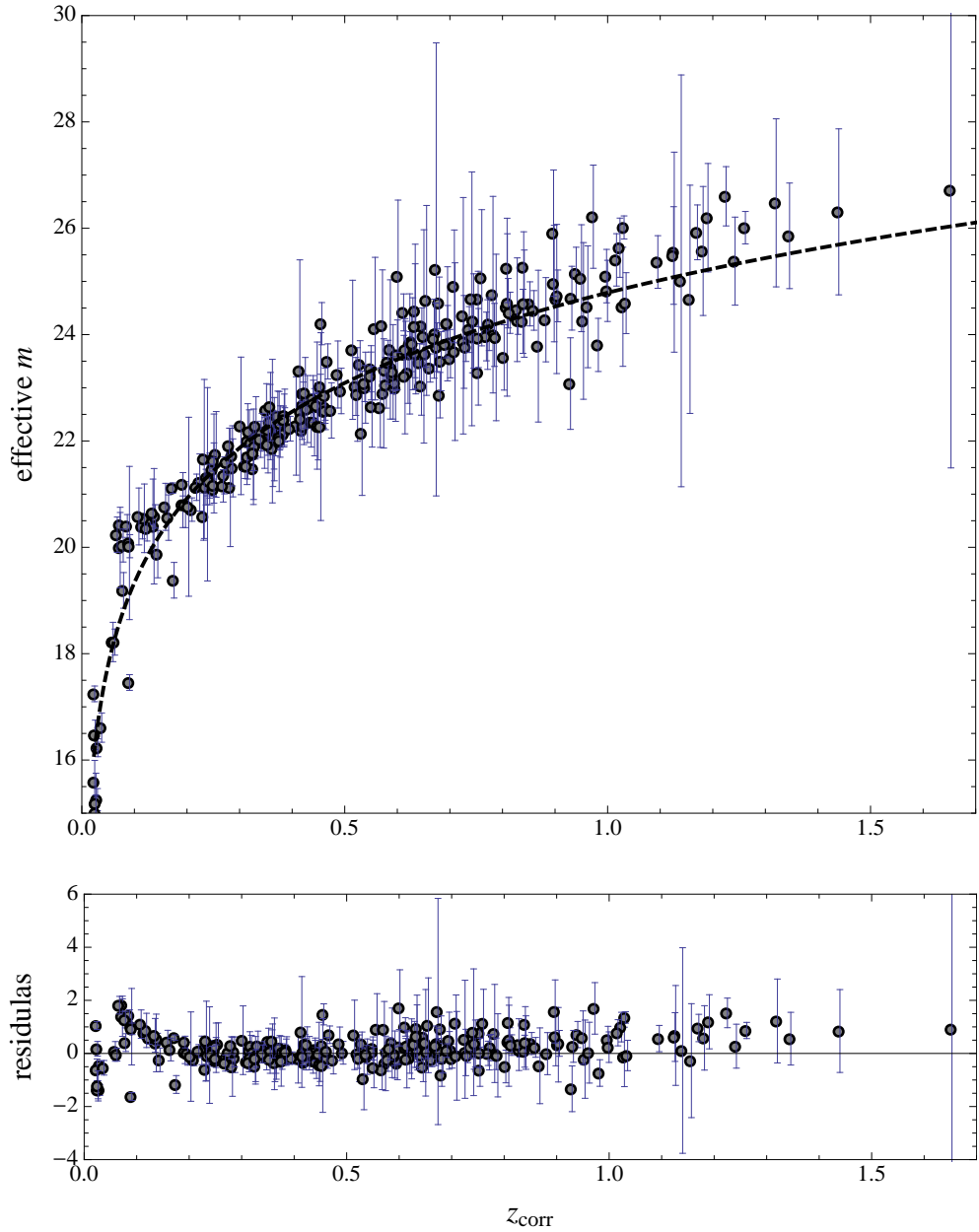}
\caption{Top: Hubble diagram for 252 type Ia SNe selected (see
text) from the ``Union compilation'' reported in
\cite{kowalski2008}. Bottom: Magnitude residuals from the
best-fit flat cosmology, ($\Omega_{\mathrm{m}}=0.70$, $\Omega_{\mathrm{\Lambda}}=0.30$).}%
\label{Fig_01}
\end{center}
\end{figure}

\begin{figure}%
\begin{center}
\includegraphics{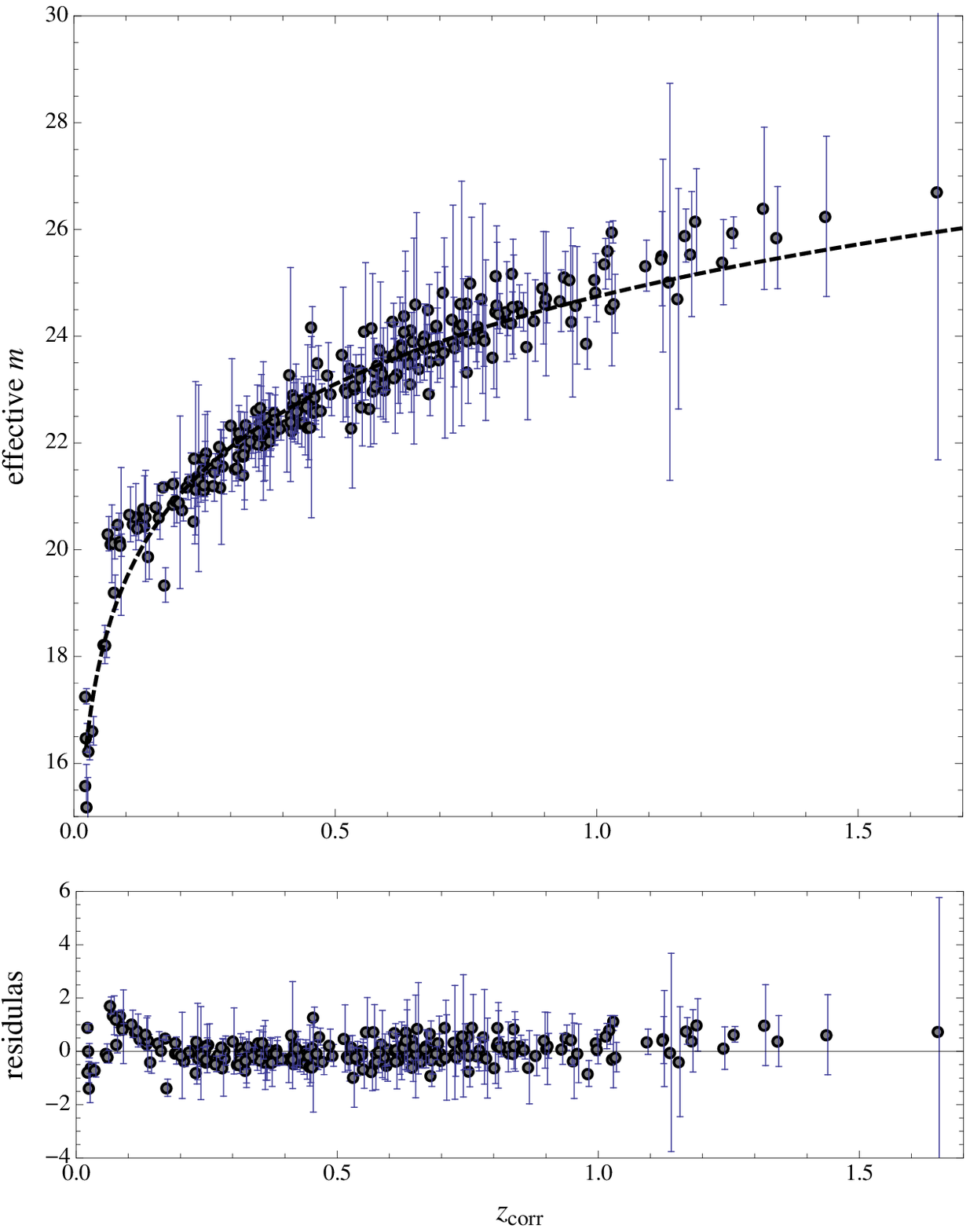}
\caption{Top: Hubble diagram for 242 type Ia SNe selected (see
text) from the ``Union compilation'' reported in
\cite{kowalski2008}. Bottom: Magnitude residuals from the
best-fit flat cosmology, ($\Omega_{\mathrm{m}}=1$, $\Omega_{\mathrm{\Lambda}}=0$).}%
\label{Fig_02}
\end{center}
\end{figure}

In order to verify the stability of our result, we removed other
10 SNe from our sample (1992bc, 2001cz, 1995ac, 1995ao, 2002w,
04D3df, 03D1cm, bo10, g055, m138). This time, the value of
$\Omega_{\mathrm{m}}$ changed again and became equal to the unity,
causing the vanishing of any form of dark energy. Note that the
minimum of $\chi^2$ in the last case falls outside the range of
the allowed values $0 \leq \Omega_{\mathrm{m}} \leq 1$. So we have
computed some three parameters fits for fixed values of $
\Omega_\mathrm{m}$ in the right range. The result is that $\chi^2$
is a decreasing function of $\Omega_\mathrm{m}$ that has the
smallest acceptable value just when we put $\Omega_\mathrm{m} =1$.
The values of the remaining three parameters are reported in Table
\ref{Table_1}. The behavior of the Hubble diagrams are shown in
figures \ref{Fig_01} and \ref{Fig_02} for the 252 and the 242 SN
subset, respectively.

%

By inspection of Table \ref{Table_1}, it is evident that the
absolute magnitude and the parameter $\beta$ are stable for the
three samples. Instead, the cosmological parameter
$\Omega_{\mathrm{m}}$ changes drastically, and also the parameter
$\alpha$ derived using $z_{\mathrm{corr}}$ is very different from
the one obtained with the standard $z$. The reason of the latter
difference can be explained expanding ``$D (z_{\mathrm{corr}})$''
in the equation (5) around $s = 1$. At the first order in $(s-1)$
we have
\begin{equation}
5 \log_{10} D(z_{\mathrm{corr}}, \Omega_{\mathrm{m}}) \simeq 5
\log_{10} D(z, \Omega_{\mathrm{m}}) + \delta(z,
\Omega_{\mathrm{m}}) (s-1),
\end{equation}
where
\begin{equation}
\delta = \frac{5}{\ln 10} \left(1 +
\frac{1+z}{\sqrt{\Omega_{\mathrm{m}} (1+z)^3 +
\Omega_\Lambda}\int_0^{z}
\frac{dz^\prime}{\sqrt{\Omega_{\mathrm{m}} (1+z^\prime)^3 +
\Omega_\Lambda}}} \right).
\end{equation}
If we insert this result in the equation (4) we can write $\chi^2
(z_{\mathrm{corr}})$  in terms of the normal $z$. Comparing it
with the corresponding standard $\chi^2 (z)$ we obtain
$\alpha_{\mathrm{norm}} = \alpha_{\mathrm{corr}} - \delta(z_i)$ in
the sum. So, considering $\bar{\delta}= \sum_{i=1}^N
\delta(z_i)/N$, which is the average value of $\delta$ in our
sample, the estimates of $\alpha$ listed in Table \ref{Table_1},
when we use the corrected redshift, are in effect
$\alpha_{\mathrm{corr}} \simeq \alpha_{\mathrm{norm}} +
\bar{\delta}$.

\section{Conclusions}
\label{Sec04}
We introduced a correction on the redsfhit of the SNe Ia that
takes into account the correction of the corresponding luminosity,
which is commonly performed in order to gets all the SN light
curves to match. In this framework, we reanalyzed the Union
compilation \cite{kowalski2008} finding the following results:

\begin{itemize}
\item[$\bullet$] We refitted the set of 307 type Ia SNe
by a fitting procedure based on the Newton method search for a
minimum. The constraint we obtained from supernovae on the dark
energy density is $\Omega_{\Lambda}=0.607 \pm 0.046$, which is
quite different from the result of $\Omega_{\Lambda}=0.713 \pm
0.028$ found in ref. \cite{kowalski2008}. This outcome reflects
the fact that the fit of the SNe is very sensible to the fitting
procedure and to the systematic and intrinsic errors
inserted in the $\chi^2$. %


\item[$\bullet$] We corrected the redshift of the SNe according to eq.
(\ref{Eq_02}), and neglected the SNe whose redshift turned out to
be negative. After a selection cut based on the residuals, we
reduced the original compilation to two subsets formed by 252 and
242 SNe respectively. Fitting the first subset, we found that the
values of the cosmological parameters were practically reversed
($\Omega_{\mathrm{m}}=0.70$, $\Omega_{\mathrm{\Lambda}}=0.30$). On
the other hand, from the fit of the second subset we find even
$\Omega_{\mathrm{m}}=1$, which corresponds to a Universe without
dark energy.


\item[$\bullet$] The fits of the SNe, both with $z$ and $z_{\mathrm{corr}}$, depend
on the number of SNe used too. The difference of few SNe between
two samples changes dramatically the estimates of the cosmological
parameters. The above mentioned case of the two subsets formed by
252 and 242 SNe respectively is emblematic.
\end{itemize}

In conclusion, we found that the constraint on the cosmological
parameters based on the analysis of SNe is contingent on both
fitting procedure and the number of SNe considered in the sample.
On the contrary, the absolute magnitudes of SNe are not affected
by variations in the sample.

We emphasize that it is well known that the hypothesis according
to which the Universe is accelerating comes not only from the
results of SNe Ia, but also from the analysis of CMB and galaxy
clusters. The latter two methods converge on a result of
$\Omega_{\mathrm{\Lambda}} \simeq 0.70$, $\Omega_{\mathrm{m}}
\simeq 0.30$. The present work is not intended to replace the
current paradigm of the cosmology (even if the debate is still
alive \cite{sarkar2008}), but only to point out that the use of
SNe Ia in the estimate of cosmological parameters can introduce,
through the stretch factor, some instability in the final result,
in addition to biases and discrepancies in the SN Ia data samples.
So, all the results that come from the SNe Ia must be taken with
more caution, especially when they are used to discriminate
between different cosmological models.

%

\ack
The data of the original \emph{Union compilation} has been
downloaded from http://supernova.lbl.gov /Union. Our numerical
analysis has benefited from the Mathematica files of S. Nesseris
and L. Perivolaropoulos, which are available on
http://leandros.physics.uoi.gr/snls.htm. A.F. and L.M. acknowledge
support for this work by research funds of the University of
Sannio.

\end{document}